\documentclass[11pt]{article}
\usepackage{epsfig,graphicx,psfrag,axodraw,bm,amssymb}
\usepackage{mathrsfs,amsfonts,hepunits}
\usepackage[hyphens]{url}\usepackage{hyperref}
\hypersetup{breaklinks=true, pagecolor=white, colorlinks=false}
\newcommand{\bear}{\begin{eqnarray}}
\newcommand{\eear}{\end{eqnarray}} 
\def\lsim{\mathrel{\rlap{\lower4.4pt\hbox{\hskip0.2pt$\sim$}}
    \raise1pt\hbox{$<$}}}
\def\gsim{\mathrel{\rlap{\lower4.5pt\hbox{\hskip0.8pt$\sim$}}
    \raise1pt\hbox{$>$}}}

\newcommand{\be}{\begin{eqnarray}}
\newcommand{\ee}{\end{eqnarray}}

\begin{document}

\twocolumn[ 

{\scriptsize FERMILAB-PUB-11-405-T} \\  

\vspace{-1.1cm}

\title{\bf \Large Very Light Axigluons and the Top Asymmetry}
\author{\large Gordan Z. Krnjaic$^{1,2}$ \\ [4mm]  
\it\small  $1)$ Theoretical Physics Department, Fermilab, Batavia, IL 60510, USA\\ [2mm]  
\it\small $2)$ Department of Physics and Astronomy, Johns Hopkins University, 
Baltimore, MD 21218, USA
}

\date{\normalsize \today}
\maketitle
 

\vspace{-.8cm}

\begin{quote}
We show that very light (50 -- 90 GeV) axigluons with flavor-universal couplings of order  $g_{s}/3$
 may explain the anomalous top forward-backward asymmetry reported by both CDF and D0 collaborations.
 The model is naturally consistent with the observed $t \bar t$ invariant mass distribution 
 and evades bounds from light Higgs searches,  LEP event shapes, and 
  hadronic observables at the $Z$ pole. Very light axigluons can 
 appear as resonances in multijet events, but searches require sensitivity to masses below current limits. 
\end{quote}

\vspace{.4cm}
]


\section{Introduction}

The CDF and D0 collaborations have recently reported 
measurements of the forward-backward asymmetry ($A_{FB}$) in
$t\bar t$ production with 
intriguing deviations from the standard model prediction.
CDF's result \cite{Aaltonen:2011kc} in the  
lepton plus jets channel 
reports an inclusive parton level asymmetry 
\be
A_{FB}  \,({\rm CDF})_{\ell j} = (15.8 \pm 7.4)\%~~.
\label{CDF}
\ee
If their measurement in the dilepton channel \cite{dilepton} is combined with this 
result, the asymmetry becomes 
\be
A_{FB}  \,({\rm CDF})_{\ell \ell + \ell j} =  (20.9 \pm 6.6)\% ~~,
\label{CDFcomb}
\ee
 and exceeds 
the standard model prediction $\simeq 5\%$ \cite{Kuhn:1998kw}-\cite{Bowen:2005ap}
by more than 2 standard deviations.

D0 performs a similar search \cite{Collaboration:2011rq} in the lepton plus jets channel and 
reports an inclusive parton-level asymmetry 
\be
A_{FB} \,({\rm D0})_{\ell j} = (19.6 \pm 6.5)\% ~~,
\label{D0}
\ee
which is also more than 2$\sigma$ above the SM result. Taken together, 
these consistent deviations may be evidence for new physics in top quark production. 

While all the inclusive measurements are consistent with each other, 
CDF's lepton plus jets search sees sharp mass dependence \cite{Aaltonen:2011kc} in
 the binned result  
\be
A_{FB}(M_{t \bar t} < 450 {\, \rm GeV}) &=& (-11.6 \pm 14.6) \% \nonumber ~~,\\
A_{FB}(M_{t \bar t} > 450 {\, \rm GeV}) &=& (47.5 \pm 11.4) \% \nonumber~~,
\ee
where the high mass bin is $3.4\, \sigma$ above the SM prediction. Neither D0 nor the complementary CDF dilepton search see the same effect; both find 
consistently positive $>2 \sigma$ deviations from the SM over the full $M_{t \bar t}$ range. 
 
It has been observed that massive gluons with axial couplings can induce a large
forward-backward asymmetry in $t \bar t$ production
by interfering with standard model processes  \cite{  Antunano:2007da}-\cite{Alvarez:2011hi}. Motivated primarily by the mass dependent CDF result, these 
models predict asymmetries that rise uniformly with invariant mass and feature a sign flip near $M_{t \bar t} \approx $ 450 GeV.
 Large (TeV scale) masses  are typically required to satisfy dijet-resonance search bounds and suppress contributions to the $t \bar t$ 
 invariant mass distribution. To produce an asymmetry with the observed sign, most models also require flavor violation and are 
severely constrained \cite{Chivukula:2010fk} by limits on flavor changing neutral currents.
For a comparison of heavy axigluons and other models that address the top asymmetry, see \cite{Gresham:2011pa}.

 Relatively lighter axigluons (400 -- 450 GeV) \cite{Tavares:2011zg} can produce a large
 top asymmetry without flavor violation, but this mass scale is in tension with dijet resonance bounds and the differential $M_{t \bar t}$ distribution. Extra field
 content is generally required to broaden decay widths and avoid resonant enhancements to top quark observables.
 
In this paper we propose a {\it very light} (50 -- 90 GeV), weakly coupled axigluon to explain the top asymmetry. The model
inherits many of the features heavier axigluons enjoy, but counterintuitively avoids their experimental
constraints by being light:  dijet resonance searches suffer from large QCD backgrounds at low invariant masses, particles below the 
$2m_{t}$ threshold do not produce bumps in the $t \bar t$ invariant mass distribution, and nonresonant production suppresses
new physics contributions to the $t\bar t$ cross section, which start at fourth order in the axigluon coupling.
We find that the strongest upper bounds in this mass range come from Tevatron searches for light Higgs bosons produced in 
association with an additional $b$-jet. The strongest lower bounds come from
UA2 dijet searches and LEP measurements of the hadronic $Z$ width.

In Section \ref{sec:model} we describe our model; in Section \ref{sec:simulation} we discuss the details of our numerical simulation; in 
Section \ref{sec:bounds} we 
address the experimental constraints; in Section \ref{sec:Afb} we compute the $t\bar t$ forward-backward asymmetry
and compare theoretical predictions with production-level data; in Section \ref{sec:conclusion} we make some concluding remarks.


\section{Model Description} 
\label{sec:model}
\setcounter{equation}{0}

\begin{figure}[t]
\vspace*{0.1cm}

\unitlength=0.7 pt
\SetScale{1.2}
\SetWidth{0.9}      
\normalsize    
{} \allowbreak

\begin{picture}(-50,50)(100,70)
\Text(195,60)[c]{\small $q$}
\Text(195,110)[c]{\small $\bar q$}
\ArrowLine(150,50)(120,62)\Gluon(150,50)(195,50){2.5}{6}
\ArrowLine(120,38)(150,50)
\Text(300,65)[c]{\small $G^\prime$}
\ArrowLine(195,50)(230,62)
\ArrowLine(230,38)(195,50)
\Text(405,60)[c]{\small $\bar t$}
\Text(405,110)[c]{\small $ t$}
\end{picture}

\vspace*{.1 cm}
\caption{Axigluon contribution to $t \bar t$ pair production. 
Interference with the standard model gluon exchange diagram 
generates  ${\cal A}^{G^{\prime}}_{int}$. }
\label{fig:axigluon-diagram}
\end{figure}
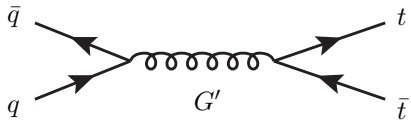

We give the axigluon ($G^{\prime}$) flavor universal couplings to SM quarks
\be 
 {\cal L} \supset  g^{\prime} G^{\prime a}_{\mu}  \bar Q  \,T^{a} \gamma^{\mu} \gamma^{5}Q ~~,
\label{eq:lag}
\ee
where $g^{\prime} \equiv \lambda g_{s}$ is the axigluon coupling constant, which we express in units of the strong
coupling. This operator can arise from an extended $SU(3)_{1}\times SU(3)_{2}$ color group that breaks 
down to the diagonal $SU(3)_{c}$ of QCD and gives rise to massive spin-1 color octets \cite{Hall:1985wz}-\cite{Hill:1991at}.
For an axigluon of mass $m_{G^{\prime}}$ our effective model requires a UV 
completion at the scale $4\pi m_{G^{\prime}} / g^{\prime} = 1.7 \, {\rm TeV}$ and 850 GeV
for $\lambda = 0.3$ and 0.6 respectively.  
In this paper, we will focus only on the low energy effective theory
and leave UV model building for future work.

Without additional field content, all decays proceed through  operator in Eq. (\ref{eq:lag}), so axigluons
can only decay to quark pairs and give rise to dijet and four jet events for single and pair production, respectively. 
 Since we work in the regime where the axigluon is below the $t \bar t $ threshold,  
 the total width is \cite{Ferrario:2008wm}
\be
\Gamma_{G^{\prime}}  = \frac{n_{f}}{6} \alpha_{s} \lambda^{2} m_{G^{\prime}} ~~,
\ee
where $n_{f}$ is the number of active fermion flavors. 
For $m_{G^{\prime}} = 80\,\GeV$ and $  \,\lambda = 0.4$, this width is $\Gamma_{G^{'}}\simeq$ 1.1 GeV.

The differential cross section for 
the process $q\bar q \to t \bar t$ in the CM frame is a sum of 
 standard model, interference, and axigluon terms
\begin{equation}
\frac{d\hat{\sigma}(G^{\prime})}{d\cos\theta}=
\mathcal{A}_{SM}+\mathcal{A}_{int}^{G^{\prime}}+\mathcal{A}_{axi}^{G^{\prime}} ~~,
\label{eq:dsdz}
\end{equation}
where  \cite{Cao:2010zb}
\be
\!\!\!\!\!\!\!\!\!\!\!\!\!\!\!  \mathcal{A}_{SM} & = &
 \frac{\pi \alpha_{s}^{2} \beta}{9\hat{s}}\left(2-\beta^{2}
 +\left(\beta\cos\theta\right)^{2}\right)~~, \label{eq:dsdz_sm} \\   \nonumber \\
\!\!\!\!\!\!\!\!\!\!\!\!\!\!\!   \mathcal{A}_{int}^{G^{\prime}} & = & \frac{4 \pi  \alpha_{s}^{2} \lambda^{2}  }{9}
 \frac{   \left(\hat{s}-m_{G^{\prime}}^{2}\right)   \beta^{2}  \cos\theta   }
{ \left(\hat{s}-m_{G^{\prime}}^{2}\right)^{2}
+m_{G^{\prime}}^{2}\Gamma_{G^{\prime}}^{2} } ~~,\label{eq:dsdz_int}   \\  \nonumber \\
\!\!\!\!\!\!\!\!\!\!\!\!\!\!\!   \mathcal{A}_{axi}^{G^{\prime}} & = &
 \frac{\pi \alpha_{s}^{2}\lambda^{4}  }{9}
 \frac{  \hat{s}  \,  \beta^{3} (1 + \cos^{2}\theta) }{\left(\hat{s}-m_{G^{\prime}}^{2}\right)^{2}
 +m_{G^{\prime}}^{2}\Gamma_{G^{\prime}}^{2}}  ~~.  \label{eq:dsdz_np}
\ee
Here $\beta \equiv \sqrt{1 - 4 m_{t}^{2} / \hat s}$ is the top quark velocity
and $ \theta$ is the angle between the incoming quark and outgoing top in the CM frame. 
A forward-backward asymmetry can only arise from terms with odd powers of $\cos \theta$, so the effect is due entirely to 
interference. In the presence of both vector and axial-vector couplings, there is an additional small contribution
 to the asymmetry from the new-physics squared term.

Note that the asymmetry generating term ${\cal A}^{G^{\prime}}_{int}$ is proportional to $(\hat s - m_{G^{\prime}}^{2})$.
For heavier axigluons, this dependence gives rise to a negative asymmetry because the mass is 
typically larger than the partonic CM energy. To compensate, many models
 introduce opposite sign couplings to the first and third generations. In our case,
$m_{G^{\prime}} < \hat s$ for on-shell $t\bar t$ production, so the asymmetry is
always positive and flavor violation is unnecessary.


\section{Simulation and Acceptances}
\label{sec:simulation}
\setcounter{equation}{0}

In the lepton plus jets analysis, CDF unfolds raw data by deconvolving their
 detector simulation and jet algorithm to yield a partonic data set 
 from events that survive cuts at the detector level. To compare our model predictions 
  with this data, it is necessary to generate an event sample with partonic $t\bar t$ pairs in the final state. However,
  knowing the predicted cross section and experimental luminosity is not enough to properly normalize
  kinematic distributions from the partonic simulation; we must also know 
  the detector level acceptances. We thus  perform two simulations:
  one at the partonic level to make our plots and one at the detector level with CDF's cuts 
  to compute the acceptances that normalize these distributions. 
 
 We simulate the partonic process $p \bar p \to t \bar t$ in
  MadGraph 5 \cite{Alwall:2011uj} using a model file  
 generated with FeynRules \cite{Christensen:2008py}.   
This file adds the operator in Eq. (\ref{eq:lag}) to the full standard 
model Lagrangian so that the 
 process in Figure \ref{fig:axigluon-diagram} contributes to $t\bar t$ production
 and gives rise to interference with SM gluon-exchange.
 
For the acceptances, we also perform a more realistic simulation
($p \bar p \to t\bar t \to \ell \nu + 4 j$)
using {\sc Pythia} \cite{pythia} for the
  parton shower and PGS \cite{pgs} for detector effects. 
 To compare with CDF's lepton plus jets search, we impose the following cuts:  at least four jets 
 with $E_{T} > 20\, \GeV$ and at least one $b$-tag;  for non-$b$ jets $|\eta_{j}| < 2$, for $b$-jets $|\eta_{bj}| <1$; 
 large missing energy $\displaystyle{\not}E_{T} > 20 \, \GeV$; and exactly one 
 electron or muon with $p_{T}^{\ell} > 20 \, \GeV$ and $|\eta_{\ell}| < 1$. 
 
 Note that there is some error introduced by this approximate method. A complete 
 comparison with experimental data would not only run a full detector simulation (including {\sc Pythia} and 
 PGS), but also identify top quarks with a least-squares kinematic fit and unfold the detector-level 
  output using the CDF algorithm that reconstructs partonic events 
   from raw data. Nonetheless, our approach accurately reproduces CDF's standard model expectation 
 for the $t \bar t$ invariant mass 
 distribution\footnote{ Although the forward-backward asymmetry arises only at loop level in the
  SM, its numerical value is tiny ($\sim 5\%$), so this tree level method 
 also adequately reproduces the (nearly symmetric) SM predictions for the $\Delta y = y_{t} - y_{\bar t}$ 
 rapidity distributions in \cite{Aaltonen:2011kc}.} 
 so the error introduced by a constant acceptance function is likely to be small in our case as well.  We leave the full unfolding
 for future work.


\section{Experimental Constraints}
\label{sec:bounds}
\setcounter{equation}{0}

Models that explain the top asymmetry must agree 
with the $t \bar t$ invariant mass distribution and total cross section, both of which are in good 
agreement with standard model predictions. Any candidate model with an $s$-channel 
mediator must satisfy constraints from dijet resonance searches at hadron colliders.
In our case, we must also contend with a variety of older measurements that set lower bounds on new 
colored particles.


\subsection{Top Quark Measurements}

The $t\bar t$ cross section at the Tevatron has been measured to be $\sigma_{t\bar t}^{\rm exp.} = 7.50 \pm 0.48 $ pb  \cite{obscx}, which 
agrees with the standard model prediction in perturbative QCD\footnote{ For complementary calculations 
see \cite{Kidonakis:2008mu, Cacciari:2008zb}. }, $\sigma_{t\bar t}^{\rm sm} \simeq  (6.32 - 7.99)$ pb for $m_{t} = 172 \,\GeV$ \cite{Moch:2008qy}. 
The leading order result, $(\sigma_{t\bar t}^{\rm sm})_{\rm LO} \simeq 5.63  $ pb, computed with MadGraph, implies a SM $K$-factor between 1.12 and 1.42.

Including an axigluon with $m_{G^{\prime}} = 80$ GeV and  $\lambda = 0.4$, gives a total LO cross section
 of $(\sigma_{t\bar t}^{\rm axi})_{\rm LO} = 6.08$ pb, which is only an 8\% increase over the SM LO result.
 This minor enhancement is due entirely to ${\cal A}^{G^{\prime}}_{axi}$ in Eq.(\ref{eq:lag}), which is
 fourth order in the axigluon coupling; the
 interference term ${\cal A}^{G^{\prime}}_{int}$ does not contribute to the total cross section. 
Although computing higher order corrections is beyond the scope of this work, the color structure of the 
axigluon exchange diagrams is identical to that of the relevant SM processes, so we expect higher order corrections 
to be of similar magnitude, though a more precise calculation is necessary to take into account the additional interference. 
 As long as the $K$ factor does not differ substantially from that of SM production, the 
total $t\bar t$ cross section stays in good agreement with experiment.
 For the remainder of this paper, we will assume the $K$ factor to be 1.2, so our benchmark 
cross section becomes 7.3 pb.

\begin{figure}[t]
\begin{center}%
\vspace{0cm}
\hspace*{-3.2cm}
\parbox{2.2in}{\includegraphics[width=.48 \textwidth]{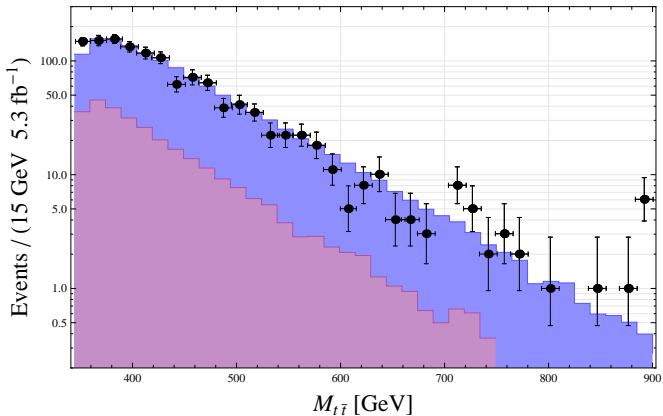}}
\vspace{-0.cm}
\caption{Tevatron invariant mass distribution for $t \bar t$ pairs (blue, color online) including both axigluon and
background contributions. Data points and standard 
model background (purple) are taken from CDF's lepton plus jets search \cite{Aaltonen:2011kc}.
Here we use  $\lambda = 0.4$ and $m_{G^{\prime}} = 80\, \GeV$.   After including a
a $K$-factor of 1.2, the top cross section is $\sigma_{t \bar t} = 7.3$ pb.  Applying the CDF cuts
 (see Section \ref{sec:simulation}) gives an acceptance of 2.6\%. 
\label{fig:invmass} }
\end{center}%
\vspace{-0.3cm}
\end{figure}

For very light axigluons $(m_{G^{\prime}} \ll 2 m_{t})$, top pair production 
is nonresonant, so the invariant mass distribution is also in good agreement with
experiment. In Figure \ref{fig:invmass} we show the simulated $M_{t\bar t}$ distribution (blue) plotted alongside the CDF data 
points and standard model background (purple) taken from the lepton plus jets search \cite{Aaltonen:2011kc}.


\subsection{Dijet Resonance Searches}
\label{sec:dijet}
Quark coupled axigluons give rise to two and four jet events from single and pair production, respectively. 
Our mass range of interest (50 -- 90 GeV) is safe from 
Tevatron \cite{Aaltonen:2008dn,Abazov:2003tj} and LHC \cite{Aad:2011aj, Khachatryan:2010jd} dijet resonance 
searches, which do not set bounds on masses below 180 and 200  GeV, respectively. 
A preliminary ATLAS analysis of multijet events \cite{newATLAS} sets limits on
color octet scalars with narrow widths, but does not constraint masses below 100 GeV. 
With lower search thresholds, this model may
be testable at both the Tevatron and LHC, however, signal and background are expected to be 
large at both colliders \cite{Dobrescu:2007xf}.

The UA2 search for hadronic $W$ and $Z$ decays \cite{Alitti:1990kw} measures the exclusive two-jet mass spectrum
between 48 and 300 GeV, which constrains the light axigluon parameter space. Using 4.7 ${\rm pb}^{-1}$ for $M_{jj} >$ 66 GeV
(and 0.58 pb$^{-1}$ for 48 GeV $< M_{jj} <$ 66  GeV), the combined $W$ and $Z$ resonances are 
extracted with a bi-gaussian fit above a smooth background function 
normalized to the data. The best fit bi-gaussian signal spans the $M_{jj}$ range between 70 and 100 GeV and 
yields a cross section of $\sigma \cdot {\cal B}r(W, Z$ $ \to jj)_{\rm obs.} = 9.6 \pm 2.3 \pm 1.1$ nb, whose
central value exceeds the SM prediction at NLO, $\sigma \cdot {\cal B}r(W, Z \to jj)_{SM} = 5.8$ nb, by almost a factor of two.

\begin{figure}[h!]
\begin{center}%
\vspace{0cm}
\hspace*{-3.2cm}
\parbox{2.2in}{\includegraphics[width=.48 \textwidth]{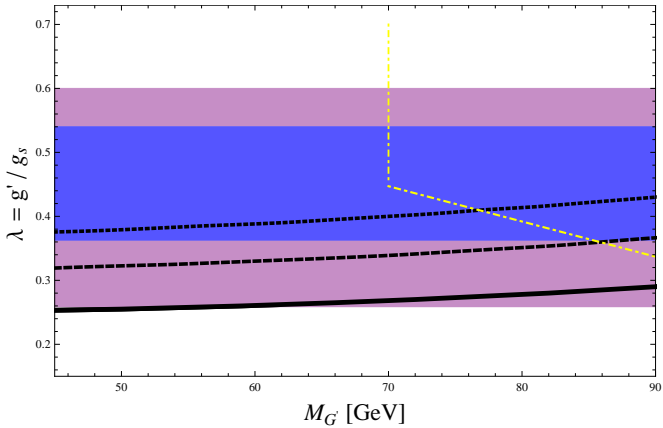}}
\vspace{-0.cm}
\caption{    Allowed axigluon  parameter space in the ($\lambda, m_{G^{\prime}})$ plane plotted alongside
bounds from dijet-resonance searches and $\Gamma({Z\to {\rm hadrons}})$ measurements
assuming different extractions of $\alpha_{s}$. 
The blue and purple bands (color online) are regions favored by the combined CDF/D0 inclusive asymmetry 
measurements at $1\sigma$ and $2\sigma$, respectively.
The dot-dashed yellow curve marks the approximate 2$\sigma$ bound above which model predictions  
exceed UA2 dijet limits from hadronic $W$ and $Z$ decays (see Section \ref{sec:dijet}).  
 The solid black curve marks the boundary above which corrections to the hadronic $Z$ 
width exceed the observed value by $2\sigma$ assuming the standard model 
extraction of $\alpha_{s}(m_{Z}) =  0.1184$. The dashed and dotted black curves
 give the same bound, but respectively assume $2.5\%$ and $5\%$ reductions 
 to the SM value of $\alpha_{s}(m_{Z})$.
Reductions of this magnitude are typical of light 
axigluon contributions to the QCD beta function (for a discussion 
see Sections \ref{sec:alpha-running} and \ref{Zwidth}).
The region above $m_{G^{\prime}} > 90 $ GeV is excluded by Tevatron $3 b$-searches.  
Since LEP event shapes rule out gluon-coupled adjoint fermions around 50 GeV, 
our model may encounter a stronger lower bound since axigluons
also couple to quarks, but a proper
analysis is necessary to set the correct limit.
 \label{fig:param} }
\end{center}%
\vspace{-0.3cm}
\end{figure}

 Although a three-gaussian fit and a QCD background prediction are necessary to properly constrain axigluons
 using this data, we can extract a rough bound by finding $(\lambda, m_{G^{\prime}})$ values
for which the combined SM and new-physics predictions exceed the observed number of events under the best fit 
gaussian by 2$\sigma$. In Figure \ref{fig:param} we plot the exclusion boundary (yellow dot-dashed line) 
determined using Madgraph, {\sc Pythia}, and PGS to simulate our signal. 

For dijet masses below 70 GeV, the UA2 analysis does not attempt to fit any signal, so a possible resonance would almost certainly have been missed
given the very large background in this mass range. Even near $m_{W}$ and $m_{Z}$, the signal/background ratio
is only a few percent and the gauge boson peak is not visible to the naked eye (see Figure 5 in \cite{Alitti:1990kw}) prior to a rescaling 
that emphasizes the region around the known $W$ and $Z$ masses. Since the background model for this search is purely data-driven, the low-mass region does
not impose a meaningful constraint without a dedicated bump hunt. 

\subsection{Light Higgs Searches}
 Tevatron searches that look for light Higgs bosons produced 
in association with $b$-jets ($p \bar p \to h b \to b b b$) are sensitive to axigluon decays into $b$-quarks. Since these searches require at least three $b$-tags
to reduce the QCD multijet background, the bounds they impose on $\sigma(h b) \cdot {\cal B}r(h\to b b)$ also apply
to the processes $p \bar p \to G^{\prime} b \to bbb$ and $p \bar p \to G^{\prime} bb \to bbbb$, the latter of 
which can also arise from pair produced axigluons. However, the CDF \cite{CDFhiggs} and D0 \cite{D0higgs} results
only apply to masses above 90 GeV; light axigluons fall below the sensitivity threshold.
  To be conservative, we will only consider masses below 90 GeV where the 
  3$b$ constraints do not apply. 

The authors in  \cite{Doncheski:1998ny} use Tevatron Higgs searches in the associated production channel,
$p\bar p \to W h \to (\ell \nu)(b\bar b)$ to exclude axigluons with $\lambda=1$
 between $75 - 125$ GeV assuming ${\cal B}r(G^{\prime} \to b\bar b) = 1/5$. In our case with $\lambda = 0.4$,
 the Tevatron $q\bar q \to W G^{\prime}$ cross section decreases by a factor of $\lambda^{2}$, 
 which reduces the axigluon signal
  $\sigma\cdot {\cal B}r$  from $\approx 50$ pb down to $\approx 5$ pb for $m_{G^{\prime}} = 50$ GeV also
  assuming ${\cal B}r(G^{\prime} \to b\bar b) = 1/5$. 
  This falls safely below the quoted bound of  $\lsim 20$ pb, however,  
  this number is based on analysis from an unpublished talk, so its status is not clear. 
 Current Tevatron searches for the associated
  production of Higgs bosons are not sensitive to masses below 
  100 GeV \cite{CDFassoc, D0assoc}. 

Naively it would appear that LEP searches in the Higgstrahlung channel \cite{a_old}-\cite{d_old} $e^{+} e^{-} \to Zh \to 4j$ would
 be sensitive to light axigluons produced in $e^{+}e^{-} \to Z^{*} \to q\bar q G^{\prime} \to 4j$ events. However,
the event selection algorithms in these analyses look for kinematics that fit the 
Higgstrahlung topology in which the invariant masses of jet pairs produce both $Z$ and Higgs
resonances. In events with on-shell axigluons, all four jets arise from virtual $Z$ exchange, so this possibility is highly 
disfavored.  Furthermore, this process occurs at 
 order $\lambda^{2}$ and suffers additional phase-space suppression.
  
 Similar considerations apply to LEP measurements of triple gauge boson couplings \cite{:2007pq}-\cite{Barklow:2001ge} 
 which look for $e^{+}e^{-} \to W^{+}W^{-},$ $ ZZ \to 4j$ events. These analyses select events
 using neural network algorithms designed to identify diboson production; light axigluons 
 arising from $Z$ exchange have very different kinematics and fail this 
 selection, which requires some combination of jet pairs to reconstruct
 at least one gauge boson mass. At the higher end of our mass range ($m_{G^{\prime}} > 80$ GeV) it
  may be possible for an axigluon to fake a hadronically decaying SM gauge boson, but the other two jets 
 would not reconstruct a resonance. The coupling and phase-space suppression also diminish the rate at 
these searches, so axigluon production is negligible compared to tree-level diboson and QCD background
processes.


\subsection{Event Shapes}
\label{sec:scet}
 Constraints on light colored-particles have been   
 extracted from the analysis of event shapes at LEP. Comparing multijet data with
 calculations in soft colinear  effective theory (SCET) rules out 
 color adjoint fermions below 51 GeV at 95\% confidence \cite{Kaplan:2008pt}. 
 However, this approach assumes that the new field couples only to gluons, with
 no tree-level quark interactions. To set a proper lower bound, it is necessary
 to repeat this analysis with more general assumptions, however, it is unlikely
 that this would yield a more lenient limit so we will not consider masses
 below $\approx$ 50 GeV. 
  
 LEP studies of four-jet events from $Z$ decays \cite{Adeva:1991dx}-\cite{Heister:2002tq} can be sensitive to
 light, colored particles that couple to quarks. Various angular distributions are used
 to successfully distinguish $SU(3)_{c}$ QCD from alternative abelian theories of the strong force, so the presence of 
 light axigluons could potentially spoil this success. However, using Madgraph to generate four-jet $Z$ decays at 
the parton level, we find that the presence of an axigluon ($\lambda = 0.4$) in our mass range
 does not qualitatively distort these angular distributions relative to the QCD prediction.
 This is unsurprising since $ {\cal O}(10\%)$ of SM hadronic $Z$ decays produce four-jets
 -- the exact number depends on $y_{cut}$ and other jet algorithm details \cite{Bethke:1990ea} -- 
  whereas  in our model  only  ${\cal O} (0.1\%)$ of hadronic decays proceed through
  $Z \to q\bar q$ $ G^{\prime}\to 4j$ prior to imposing cuts (see Section \ref{Zwidth}). 
For higher energies probed by LEP II ($\sqrt{s} \approx$ 200 GeV), the total
 $e^{+}e^{-}\!\to Z^{*}\to q\bar q G^{\prime} \to 4 j$ rate 
  is similarly negligible compared to SM four jet production; this 
  conclusion is robust for values of $y_{cut}$ spanning several orders of magnitude.


\subsection{Running of $\alpha_{s}$}
\label{sec:alpha-running}

Since axigluons couple to the strong sector, they give rise to loop diagrams 
that modify the QCD beta function above the scale $m_{G^{\prime}}$. 
The standard model running between energy scales $Q$ and $\mu$ is given by
   \be
   \alpha_{s}(Q^{2}) = \frac{\alpha_{s}(\mu^{2})}{1+ b\, \alpha_{s}(\mu^{2}) \log\left(\frac{Q^{2}}{\mu^{2}}\right)  }~~,
   \ee
where, to leading order, $b = (33 - 2 n_{f})/12 \pi$ and $n_{f}$ is the number of active
flavors. Since axigluons have the same quantum numbers and self couplings as gluons, their
principal effect on the running is to double the gluon contribution to the beta function 
above  $m_{G^{\prime}}$:  $b \to (2\times 33 - 2 n_{f})/12 \pi$. This accelerates asymptotic
freedom and yields smaller values of $\alpha_{s}$ near the weak scale. 

While this adjustment na\"{i}vely
 jeopardizes the agreement between theory and experiment for
 the running, the experimental extraction of $\alpha_{s}$ 
 depends entirely on the assumed validity of standard model QCD with no additional
 field content \cite{Bethke:2009jm}. 
At each energy scale, an $\alpha_{s}$-dependent observable is equated to the 
 SM prediction and the resulting data point is extracted implicitly. If
 light new states were present in the strong sector, this 
 data would completely ignore their contributions, so the 
 current agreement between theory and experiment 
  does not constrain our model. 
  
  To roughly estimate the axigluon correction to $\alpha_{s}(m_{Z})$, we
   use a well-measured value of $\alpha_{s}$ below $m_{G^{\prime}}$ as
   an IR  boundary condition and evolve it with the new beta
   function. This method is crude because even low-energy observables used
   to extract $\alpha_{s}$ depend somewhat on virtual axigluon processes,
   which are ignored in the 
   extraction of reported measurements. Nonetheless, using 
   the boundary condition $\alpha_{s}(14.9 \,{\rm GeV}) = 0.160$,  
   \cite{Bethke:2009jm} the weak-scale value becomes 
   $\alpha_{s}(m_{Z}) = 0.105, 0.110$, and 0.115 for $m_{G^{\prime}} = 50, 65$ and 80 GeV, respectively.
    Different IR boundary conditions give similar downward corrections of order a few percent relative to the 
   SM extraction $\alpha_{s}(m_{Z}) = 0.1184$. 
   Note that this result is independent of $\lambda$ since axigluons couple 
   to gluons with QCD strength. 
 
This model also predicts a kink in the running of $\alpha_{s}$ near 
$m_{G^{\prime}}.$ Our mass range
 of interest (50 -- 90 GeV), however, overlaps with a region where data points are sparsely 
distributed with relatively large error bars (see Figure 6 in  \cite{Bethke:2009jm}) compared to the data set as a
whole. Kinks in the slope of $\alpha_{s}$ would, therefore, be unlikely to stand out in the data.  
Nonetheless, a model-dependent extraction of $\alpha_{s}$ is necessary to  evaluate the possibility of kinks or overall data shifts due to 
new physics contributions.


\subsection{Hadronic $Z$ Width}
\label{Zwidth}

The strongest lower bound on $m_{G^{\prime}}$ comes from virtual and three-body corrections to the hadronic $Z$ width. Axigluons 
that couple to quarks with QCD strength ($\lambda = 1$) 
enhance this width by a factor of 
\be
1 + \frac{\alpha_{s}}{\pi} f \left( m_{Z} / m_{G^{\prime}}   \right) + {\cal O}(\alpha_{s}^{2}) ~~,
\label{eq:Zcorrection}
\ee
where $f$ is a function derived in \cite{Cuypers:1991fe, Cuypers:1989nr}. 
The LEP measurement of $\Gamma(Z \to {\rm hadrons})$ and the extracted value
of  $\alpha_{s}(m_{Z})$ constrain 
the size of $f(m_{Z}/m_{G^{\prime}})$ and
 severely restrict axigluon masses: $m_{G^{\prime}} > 570$ $\, (365) \, \GeV$ for $\lambda=1$ at the 65\%  (95\%)
  confidence level \cite{Doncheski:1998ny}.  

However, $f$ is highly nonlinear, so the mass constraint is {\it extremely} sensitive
 to the axigluon coupling. In our scenario, the constraint on $f$
  applies to the combination $\lambda^{2} f$, which dramatically 
 weakens the lower bound on $m_{G^{\prime}}$. Furthermore, following the 
 discussion in Section \ref{sec:alpha-running}, light axigluon  ($m_{G^{\prime}} < m_{Z}$) contributions to the QCD 
 beta function generically decrease the value of $\alpha_{s}(m_{Z})$ at the percent level.  
Since this is used to compute QCD corrections to the SM prediction for $\Gamma(Z \to {\rm hadrons})$ \cite{Chetyrkin:1996hq}, a smaller 
value opens up more allowed parameter space for new physics; the positive axigluon contribution to the
 width compensates for a slightly smaller SM result which is reduced by the new value of $\alpha_{s}.$
 
In Figure \ref{fig:param} we plot $2\sigma$ exclusion bounds from the hadronic $Z$ width 
on the $(\lambda, m_{G^{\prime}})$ plane alongside the regions
favored by combined CDF and D0   $A_{FB}$ measurements (discussed in Section \ref{sec:Afb}).
The solid black curve uses the standard model extraction $\alpha_{s}(m_{Z})  = 0.1184 \pm 0.0007$\cite{Bethke:2009jm}
and the measured $\Gamma(Z \to {\rm hadrons}) = 1.744 \pm 0.002$ GeV  \cite{Nakamura:2010zzi}
to identify parameters for which the theoretical prediction exceeds the measured central value by 2$\sigma$. Also plotted
are the 2$\sigma$ bounds assuming 2.5 \% (black dashed) and 5\% (black dotted) reductions in $\alpha_{s}(m_{Z})$ 
due to the modified running that includes axigluon contributions. These curves show how sensitive the bound is to 
modifications in $\lambda$ and $\alpha_{s}(m_{Z})$. Since we generically expect light axigluons to
reduce the value of $\alpha_{s}(m_{Z})$ by a few percent relative to the SM extraction, the dashed and dotted curves 
are more faithful to the underlying physics. Given the sensitivity of the bound,
a proper extraction of $\alpha_{s}$ involving axigluon processes is necessary
to accurately constrain the parameter space; the limits in Figure \ref{fig:param}
serve merely to illustrate the impact on the allowed region.


 \subsection{Bounds from $\sigma(e^{+} e^{-} \to {\rm hadrons})$}

The authors in \cite{Cuypers:1991fe} calculate
  \footnote{Note that  \cite{Cuypers:1991fe}  corrects some minor, 
yet consequential errors from an earlier paper \cite{Cuypers:1989nr} that
 placed a far stronger lower-bound on the mass.} axigluon corrections to the ratio
\be
R(s) \equiv \frac{\sigma(e^{+}e^{-}\!\to {\rm hadrons})}{e^{4}/12\pi s}
\ee
 at the scale $\sqrt{s} = 34$ GeV and thereby exclude masses below 50 GeV at 95\% confidence 
 assuming  $\lambda = 1$. 
 As with the hadronic $Z$ width, the corrections for this process are 
proportional to the factor in Eq.\,(\ref{eq:Zcorrection}) with the replacement
 $\alpha_{s}\to \lambda^{2}\alpha_{s}$, so the discussion in Section \ref{Zwidth} applies
 to this bound as well. 
Since $\Gamma(Z\to {\rm hadrons})$ is extracted from 
 $R$ data at the $Z$ pole,
 the allowed parameter space in Figure \ref{fig:param} is automatically
 consistent with bounds from $R$ near $\sqrt{s} = m_{Z}$. For smaller energies
 in our range of interest, $\sqrt{s} \in 50 - 90$ GeV, the uncertainties on the 
 $R$ data are larger than those at the $Z$ pole \cite{Nakamura:2010zzi}, so the
 bound is weaker.  
 
 \begin{figure}[t]
\begin{center}%
\vspace{0cm}
\hspace*{-3.2cm}
\parbox{2.2in}{\includegraphics[width=.48 \textwidth]{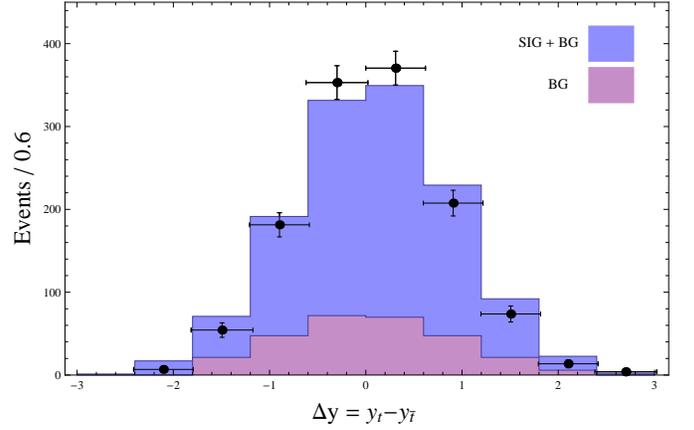}}
\vspace{-0.cm}
\caption{ Inclusive top anti-top rapidity difference distribution plotted against unfolded CDF data. 
Here we use the same model parameters as in Fig. \ref{fig:invmass}. The blue histograms 
 include both signal and standard model background. Both data and background (purple) are taken from \cite{Aaltonen:2011kc}. 
This plot omits the small, loop level asymmetry generated by SM processes.
\label{fig:LJdely} }
\end{center}%
\vspace{-0.3cm}
\end{figure}


\section{Forward Backward Asymmetry}
\label{sec:Afb}
\setcounter{equation}{0}

The forward-backward asymmetry can be written  
\be 
A_{FB} \equiv \frac{       N(\Delta y >0) -  N(\Delta y < 0)             }{     N(\Delta y >0) +  N(\Delta y < 0)   }~~,
\ee
where $\Delta y \equiv y_{t} - y_{\bar t}\, $  is the rapidity difference between the top and anti-top quarks.

In Figure \ref{fig:param} we show the favored parameter space in the $(\lambda, m_{G^{\prime}})$ plane. The blue (purple) band represents 
the region of $1\sigma$ ($2\sigma$) agreement with the combined CDF,  Eq. (\ref{CDFcomb}), 
and D0, Eq.(\ref{D0}) inclusive measurements. For typical points in these regions, 
the model predicts a positive asymmetry of order $20 \%$. 

In Figure \ref{fig:LJdely} we show the inclusive $t \bar t$ rapidity-difference distribution plotted 
against the CDF  data. The signal simulation
is identical to that used to generate Figure \ref{fig:invmass} with $m_{G^{\prime}} = 80$ GeV and $\lambda = 0.4$. After applying
the cuts described in Section \ref{sec:simulation}, the acceptance is is 2.6\%.
This plot only depicts the effects of tree-level processes; the histograms do not include the small 
asymmetry induced by standard model processes. However, the numerical results in Fig. \ref{fig:param} include the full asymmetry
with both SM and new physics contributions. 

Although our simulation gives an acceptable fit to the rapidity data, some of the bins are more than $1\sigma$ away from 
data points. We, however, do not expect perfect agreement at this level of analysis. The distribution in Figure \ref{fig:LJdely} is a 
rough approximation of the full theory prediction which requires both 
a full CDF detector simulation and the subsequent unfolding for a proper comparison with data. 

\begin{figure}[t]
\begin{center}%
\vspace{0cm}
\hspace*{-3.2cm}
\parbox{2.2in}{\includegraphics[width=.48 \textwidth]{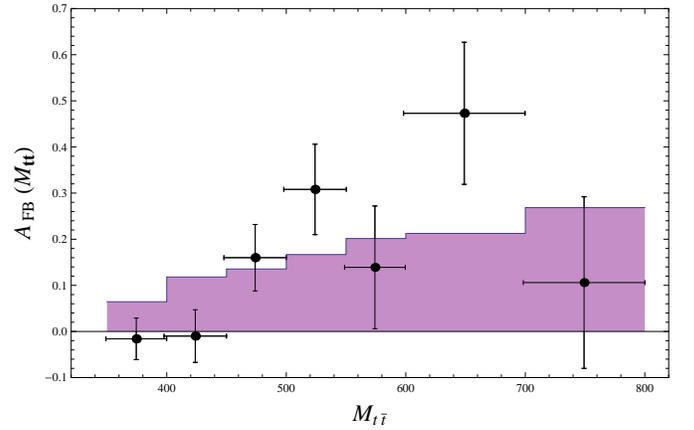}}
\vspace{-0.cm}
\caption{ Theory prediction for the mass dependent $t\bar t$ asymmetry (purple histograms)
 plotted against the binned, unfolded CDF data in the lepton plus jets channel \cite{Aaltonen:2011kc}.
Here we use the same model parameters as in Fig. \ref{fig:invmass}. 
For comparison with CDF, the bin sizes are 50 GeV for $M_{t\bar t} < 600$ GeV and 
100 GeV for larger invariant masses.  
Since the interference term in the differential
cross section,
 Eq.\,(\ref{eq:dsdz_int}), is proportional to $(\hat{s} - m_{G^{\prime}}^{2})$, 
the asymmetry is always positive for on-shell $t\bar t$ production. This is a generic feature of 
light axigluon models.
\label{fig:massdep} }
\end{center}%
\vspace{-0.3cm}
\end{figure}

In Figure \ref{fig:massdep} we show the theory prediction for the mass dependent asymmetry $A_{FB}(M_{t\bar t})$ plotted 
alongside the unfolded CDF data. Like other light $s$ channel 
 mediators, light axigluons predict a positive asymmetry throughout  the whole range of invariant masses. 
 While the agreement at low invariant mass is not ideal, neither D0 nor the CDF dilepton measurement 
 observe strong mass dependence, so the significance of the mass-dependent data is not clear. 

Note that in Figures \ref{fig:invmass}, \ref{fig:LJdely} and \ref{fig:massdep} we only compare the model to  
CDF results because their published distributions feature production-level data, which allow for a direct comparison
with parton level simulations. Comparison with D0's distributions requires a detailed understanding 
of their detector simulation, which is beyond the scope of this work. Our conclusions have 
emphasized inclusive results from both collaborations since these are in better agreement with 
each other than the more controversial mass-dependent data. 

\section{Conclusions}\setcounter{equation}{0}
\label{sec:conclusion}

We have shown that a light axigluon with flavor universal couplings 
 can generate a large, positive $t\bar t$ asymmetry and naturally agrees with
 measurements of $d\sigma /dM_{t\bar t}.$ The 
model has viable parameter space consistent with light Higgs bounds, dijet 
resonance searches and measurements of the hadronic $Z$ width.
 
For masses between $50-90$ GeV and quark couplings in the range $0.3 \,g_{s} - 0.6\,g_{s} $, the 
theoretical prediction for the parton-level top asymmetry is in good agreement with inclusive results from both CDF and D0. 
The asymmetry is proportional to $(\hat s - m_{G^{\prime}}^{2})$, so the sign of $A_{FB}$ 
is always positive 
for on shell top pair production with $\sqrt s > 2 m_{t} \gg m_{G^{\prime}}$.
 
 In the presence of a light axigluon, both the predicted and observed values of $\alpha_{s}$ 
 are modified at the percent level. A reanalysis of $\alpha_{s}(\sqrt{s})$ measurements could reveal small downward shifts in the 
 data since the modified beta function accelerates the running of $\alpha_{s}$ in the presence of an axigluon. 
 The downward shift in $\alpha_{s}$ also decreases the 
 SM predictions for $\Gamma(Z \to {\rm hadrons})$ and $\sigma(e^{+} e^{-} \to {\rm hadrons})$, which expands
 the parameter space for $(\lambda, m_{G^{\prime}})$ values that explain
 the top asymmetry. 
  
Although the QCD background at low masses is formidable, it may be possible to revisit 
UA2 dijet data and perform a dedicated bump hunt in the low mass region with 
updated background calculations.  It should also be possible to include light axigluons 
in a SCET reanalysis of event shapes in LEP data, which would likely set the strongest lower bound
 on this model.

If very light axigluons explain the top forward-backward asymmetry, 
the Tevatron and LHC experiments should, in principle, be able to observe resonances 
in two and four jet events from single and pair production. Since the 
effective model presented in this paper demands a UV completion at energy 
scales near the LHC's designed sensitivity, we predict new physics 
around the TeV scale, but the specific signals are model dependent at this 
level of description and would be interesting to pursue as future work.   

\medskip
\noindent {\bf Acknowledgments:} 
We thank Johan Alwall, Bogdan Dobrescu, David Fehling, Patrick Fox, Graham Kribs, Roni Harnik, 
David E. Kaplan, Martin Schm-\\ altz,  Daniel Stolarski, and Morris Swartz for helpful discussions. GZK is supported by a Fermilab 
Fellowship in Theoretical Physics and by the National Science Foundation under grant number 106420. Fermilab is operated by Fermi Research Alliance, LLC, under Contract DE-AC02-07-CH11359
with the US Department of Energy.

 \vfil 
\end{document}